\documentclass[preprint,twoside,superscriptaddress,showpacs,floatfix]{revtex4-2}
\usepackage{amssymb,amsmath,graphicx,multirow,natbib,hyperref,color}
\hypersetup{colorlinks,citecolor=black,filecolor=black,linkcolor=black,urlcolor=black}

\begin{document}
\title[]{Circular current induced by angular dynamics in swarmalator populations}

\author{Hyun Keun \surname{Lee}} 
\affiliation{Department of Physics and Institute for Advanced Physics,
Konkuk University, Seoul 05029, Korea}
\author{Hyunsuk \surname{Hong}}
\email{Corresponding author: hhong@jbnu.ac.kr}
\affiliation{Department of Physics and Research Institute of Physics and Chemistry, 
Jeonbuk National University, Jeonju 54896, Korea}

\date{\today}
\begin{abstract}
We propose a modified swarmalator model that generates collective rotational currents
in phase synchronization. 
Our approach builds on the original swarmalator model~\cite{Kevin17}, introducing a 
key modification: the phase-dependent terms in the spatial dynamics are replaced with a 
simpler driving term that depends on both the phase and a specified origin. 
We investigate the dynamics of this model through extensive numerical simulations. 
When the origin is fixed, spiral motions of synchronized and clustered swarmalators emerge 
from a finite fraction of random initial conditions, resulting in collective currents. 
To prevent the unrealistic divergence of these spirals, we introduce a dynamic origin, 
defined as the center of the swarmalators' positions. With this dynamic origin, the system 
evolves into rotating collective currents, where synchronized swarmalators form 
stable circular patterns. 
In both the fixed and dynamic origin cases, we also observe {\it{no-current}} states, in 
which synchronized swarmalators aggregate near the origin.
Finally, we find that the formation of collective currents can be facilitated by tuning 
the phase variables either at initialization or during the system's evolution.
\end{abstract}


\maketitle

\section{\label{intro}Introduction}
Many biological systems in nature exhibit intriguing {\it{rotational}} behaviors. For example, 
rotating ant mills and schooling fish often display circular motion patterns~\cite{circularmotion}.
There has been intense research in the last decade
concerning chiral active matter, which performs circular motion and often
displays synchronization/collective effects. Chiral active matter models the
collective behavior of animals, robots, and colloids 
that rotate~\cite{Liebchen22,Caprini24}.
In this work, we propose a modified {\it{swarmalator}} model that generates collective
rotational currents driven by synchronization among interacting agents 
such as swarmalators.

The term {\it{swarmalator}}, introduced in Ref.~\cite{Kevin17}, refers to mobile oscillators 
that couple internal phase synchronization dynamics~\cite{Kuramoto7584,vkm2-2,vkm3,Strogatz00,Pikovsky03,Acebron05,I.A14,tanaka2014,LHK23} with spatial interactions~\cite{C.W.R87,T.V95,T.V95-2,E.R95,O.J.O98,S.H04,H.K.L04,C.M.T06,gLee21}.
Previous studies on swarmalators span a range from biologically inspired, application-oriented 
models~\cite{tflk2,MS20,LA20,JM20,Lee21,tswm2,tflk-0,tflk} to more abstract mathematical 
and theoretical frameworks~\cite{Kevin18,Kevin22,Yoon22,Hong23,Kevin24}.
Understanding the collective behavior of these systems is relevant not only for 
practical applications such as the coordination of robotic or drone swarms~\cite{BBB19,tswm3}, 
but also for gaining deeper insight into the self-organizing phenomena 
observed in active matter and biological colonies~\cite{K.E07,R.B16,K.O20,A.P22,T.T22};
a comprehensive overview can be found in Ref.~\cite{Kevin19}.
Given the inherent mobility of swarmalators, the emergence of collective currents-that is, coherent, directional flows-is a natural phenomenon of interest. 
Such currents are widely observed in biological systems, including bird flocks, fish schools, 
and mammalian herds~\cite{C.W.R87,T.V95,T.V95-2,E.R95,O.J.O98,S.H04}, as well as in 
active matter and insect colonies~\cite{tflk-0,tflk}.

In this study, we focus on the numerical realization of collective rotational currents within a swarmalator system. 
In Ref.~\cite{Kevin17}, the rotating state called the {\it active phase wave} was introduced, 
but this pattern emerges from phase desynchronization via negative coupling and consists of 
pairwise symmetric microscopic movements that cancel out, resulting in zero net flow.  
While this state gives the appearance of rotating motion, it does not imply a 
genuine collective current in the physical sense.
To explore the possibility of collective flow in swarmalator system, 
we propose a modified model 
that may result in nonzero phase-induced rotational currents as emergent phenomena.
Our model preserves the phase dynamics of the original model~\cite{Kevin17}, 
but replaces the phase-dependent spatial dynamics part with a 
driving term that depends on both phase and position of the swarmalator to update.
We numerically study the model to examine the current 
including the influence by the choice of origin.

In Sec.~\ref{mm}, we briefly review how the original model leads to net-zero current, and then introduce the modified version that incorporates a phase-driven term. 
In Sec.~\ref{nr}, we present the numerical simulations that reveal two distinct types of current states: one is the outward spiral by the cluster 
composed of all swarmalators in system with fixed origin
while the other is the confined rotation of circular pattern of 
swarmalators all, in system, enclosing the center-of-mass origin.
In both cases, we also observe non-current states, where synchronized 
swarmalators gather near the origin. Our findings are summarized in Sec.~\ref{sum}.

\section{\label{mm}Motivation and model}
In this section, we first revisit the original model~\cite{Kevin17} to know it 
ultimately yields net-zero current.
Let ${\mathbf r}_i=\left(x_i(t),y_i(t)\right)$ and $\theta_i(t)$ denote 
the position and phase of swarmalator $i$ at time $t$, respectively.
The structure of the spatial dynamics in the swarmalator models~\cite{Kevin17,gLee21} is 
\begin{equation}
	\label{smbt}
	\dot{\mathbf r}_i=\frac{1}{N}\sum_j f(\|{\mathbf r}_j-{\mathbf r}_i\|,|\theta_j-\theta_i|)
	({\mathbf r}_j-{\mathbf r}_i) \,,
\end{equation}
where $N$ denotes the number of swarmalators, and $f$ is a scalar function that 
depends on spatial and phase differences. 
The concrete form of $f(\|{\mathbf r}_j-{\mathbf r}_i\|,|\theta_j-\theta_i|)$ is $A+J\cos(\theta_j-\theta_i)-B/\|{\mathbf r}_j-{\mathbf r}_i\|^2$ or $(A+J\cos(\theta_j-\theta_i))/\|{\mathbf r}_j-{\mathbf r}_i\|-B/\|{\mathbf r}_j-{\mathbf r}_i\|^2$ with parameters $A$, $J$, and $B$.
For simplicity, we use the shorthand notation 
$f_{i,j}\equiv f(\|{\mathbf r}_j-{\mathbf r}_i\|,|\theta_j-\theta_i|)$ 
throughout the paper, unless otherwise noted nor ambiguity arises.

The microscopic rotational current of swarmalator $i$ is given by 
\begin{align}
{\mathbf r}_i\times \dot{\mathbf r}_i 
= (1/N)\sum_j f_{i,j} \left({\mathbf r}_i\times {\mathbf r}_j\right)
\end{align}
for the vector/cross product $\times$. The total rotational current is then
\begin{eqnarray}
	\label{tl0}
	\sum_i {\mathbf r}_i\times \dot{\mathbf r}_i
	&=&\frac{1}{N}
	\sum_{i,j}f_{i,j} \left({\mathbf r}_i\times {\mathbf r}_j\right)
	\nonumber\\
	&=&\frac{1}{N}\sum_{i>j}f_{i,j} 
	\left({\mathbf r}_i\times {\mathbf r}_j
	+{\mathbf r}_j\times {\mathbf r}_i\right)=0,
\end{eqnarray}
where we used the symmetry of $f_{i,j}=f_{j,i}$ and the antisymmetry of the 
cross product, i.e., ${\mathbf a}\times {\mathbf b}=-{\mathbf b}\times {\mathbf a}$.
This demonstrates that the dynamics described by Eq.~\eqref{smbt} does not generate a 
collective rotational current. 
Similarly, a collective linear current is also not produced in that
\begin{align}
\sum_i \dot{\mathbf r}_i 
&= \frac{1}{N}\sum_{i,j}f_{i,j}({\mathbf r}_j-{\mathbf r}_i) \nonumber\\
&= \frac{1}{N}\sum_{i>j}f_{i,j}({\mathbf r}_j-{\mathbf r}_i
+{\mathbf r}_i-{\mathbf r}_j)=0.
\end{align} 
Therefore, to generate a collective non-zero net current, a new term must be additionally 
considered to Eq.~\eqref{smbt}.

Swarming behavior typically emerges from interactions among self-propelled agents.
In the well-known Vicsek model~\cite{T.V95}, a simple 
spatial driving is introduced to model the self-propulsion observed in bird flocking.
A similar mechanism is employed in an oscillator model
for the study of chiral active matter~\cite{tflk-0}, 
where phase variable plays the role of moving direction of particle.
Continuing the idea of Vicsek-style driving, 
we extend the spatial driving dynamics by incorporating spatial information also.
We expect that this modeling may provide 
such a driving term that can convert the linear spatial drift in phase synchronization 
into a collective rotational motion.
We remark that Refs.~\cite{tflk-0,tflk} introduce non-zero natural frequencies 
into phase dynamics as a mean to generate circular motion.

Our modification to the standard swarmalator model~\cite{Kevin17} is as follows.
The driving term sketched above replaces 
the phase-dependent part in the existing spatial dynamics. Equation~\eqref{smbt} shows
the spatial dynamics is structurally governed by
position differences. Therein, the role of phase 
is restricted to a sort of parameter-variation that 
however becomes disable in phase synchronization 
as phase-difference is actually referred to. 
Phase-sync is an interesting steady-state property of 
interacting oscillators system. Therefore, we believe such a model 
where this steady property plays a role in the long-time spatial dynamics is 
more interesting and meaningful. To this end, we remove the phase-dependent 
part from the spatial dynamics, and then separately
add a driving term whose arguments are the phase and position of the swarmalator  
to update. This way, the phase variable is structurally involved into the onset 
and maintenance of collective current (if emerges).

\begin{figure}
\includegraphics[width=6cm]{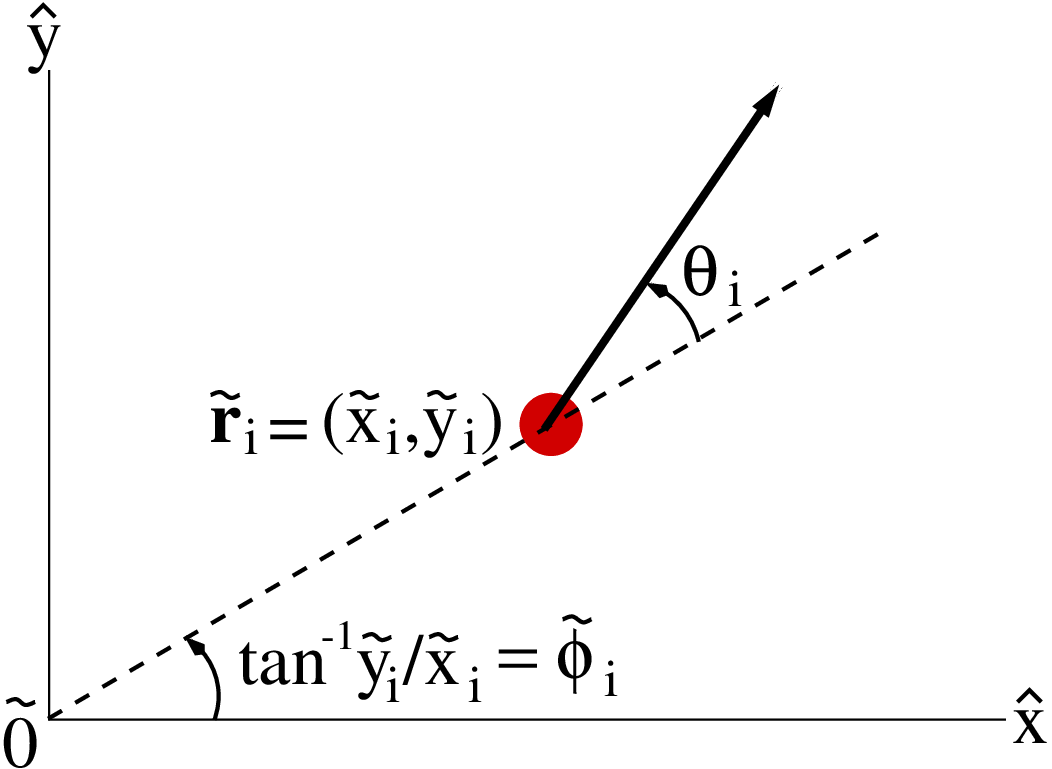}
\caption{Direction of the driving motion induced by the $W$-term in Eq.~\eqref{model}.
	Here, $\tilde{\mathbf r}_i=(\tilde x_i,\tilde y_i)$ denotes the position of 
	swarmalator $i$ in the rectangular coordinate system centered at the chosen 
	origin $\tilde O$, i.e., $\tilde{\mathbf r}_i\equiv {\mathbf r}_i-\tilde O$. 
	When $\tilde O=O$ for the fixed origin $O=(0,0)$, 
	we have 
	$\tilde{\mathbf r}_i= {\mathbf r}_i$. The direction vector $\hat x$ and $\hat y$ 
	remain fixed regardless of the choice of $\tilde O$. The arrow originating from 
	$\tilde{\mathbf r}_i$ indicates the direction of the driving by the $W$-term.
	As shown, the phase $\theta_i$ is measured counterclockwise from the radial 
	direction of $\tilde{\mathbf r}_i$.
}
\label{fg-model}
\end{figure}

The model we will study is 
\begin{eqnarray}
	\label{model}
	\dot{\mathbf r}_i 
		&=& \frac{A}{N}\sum_j \left(1
	-\frac{1}{\|{\mathbf r}_j-{\mathbf r}_i\|^2} \right)
		\left({\mathbf r}_j-{\mathbf r}_i\right)
	+W \left(\hat {\mathbf x}\cos(\tilde\phi_i+\theta_i)
	+\hat {\mathbf y}\sin(\tilde\phi_i+\theta_i)\right),
	\\
	\label{model2}
	\dot \theta_i 
		&=& \frac{K}{N}\sum_j 
	\frac{\sin(\theta_j-\theta_i)}{\|{\mathbf r}_j-{\mathbf r}_i\|}\,,
\end{eqnarray}
where $\tilde\phi_i=\tilde\phi_i(\{\tilde{\mathbf r}_i\})=\tan^{-1}(\tilde y_i/\tilde x_i)$ for ${\tilde{\mathbf r}}_i\equiv {\mathbf r}_i-\tilde O$ with
the movable origin $\tilde O = {\tilde{O}}(\{{\mathbf r}_i\})$
in the choice of $\{{\mathbf r}_i\}$-dependence.
The fixed origin $O \equiv (0,0)$ is, of course, a choice. 
We have omitted the tilde notation in the $A$-coefficient-term (hereafter referred to as the 
$A$-term), since the difference ${\tilde{\mathbf r}}_i-{\tilde{\mathbf r}}_j =
{\mathbf r}_i-{\mathbf r}_j$ holds regardless of the choice of $\tilde O$.
The unit vectors $\hat{\mathbf x}$ and $\hat{\mathbf y}$ used in the $W$-term represent 
the standard Cartesian basis and are independent of $\tilde O$. 
Figure~\ref{fg-model} shows the direction of the $W$-term, which depends on the position ${\mathbf r}_i$ of each swarmalator regardless in 
phase sync ($\theta_i=\theta_{\rm s}$ for all $i$) or not. 
This position-dependent driving is a distinctive feature of our modified model, which is not the case in the existing studies~\cite{T.V95,tflk-0}.
%%%%%%%%%%%%%%%%%%%%%%%%%%%%%%%%%
%%%%%%%%%%%%%%%%%%%%%%%%%%%%%%%%%

When the spatial angle $\tilde \phi_i$ is not taken into account 
in the driving term of Eq.~\eqref{model}, the driving becomes 
the very spatial movement rule considered in the Vicsek and chiral 
models~\cite{T.V95,tflk-0}. With phase angle only, this movement 
is simply a linear drift along the direction the phase angle is pointing. 
So that, if a rotational motion is of interest, another modeling looks necessary 
like an introduction of natural frequency as done in the chiral models. 
Instead of this rather direct 
implementation, we exploit the combined play of the spatial and
phase angles. Simply, we use their sum for the argument of the driving term.
As illustrated in Fig.~\ref{fg-model}, the phase angle $\theta_i$
considered in the presence of the spatial angle $\phi_i$
yields the perpendicular component to the position vector of swarmalator $i$.
This brings about the circular motion with respect to the origin $\tilde O$. 
Obviously, the rotation expected this way is rather restrictive and, furthermore, 
may blur due to the numerous interplay counted in the $A$-term, 
and this is not the case for such rotation by 
natural frequency in the absence of interplay~\cite{T.V95,tflk-0}.
However, it is also interesting to examine whether a non-direct 
microscopic modeling may result in an emergent collective phenomenon.
We are interested in whether the apparent rotating factor, at least, microscopically and temporally could result in a collective 
and a long-term circular current.

In the following, we numerically investigate Eqs.~\eqref{model} and \eqref{model2} 
with $K>0$ to examine whether a collective current can emerge in the synchronized phase.
We consider two cases for the origin $\tilde O$: a fixed origin at $O=(0,0)$, and 
a dynamic origin given by the position center (PC) of the swarmalators, 
defined as $\tilde O=O_{\rm PC}\equiv\sum_i {\mathbf r}_i/N$.

\section{\label{nr}Collective current in phase sync}

The dynamics of the model described by Eqs.~\eqref{model} and \eqref{model2} 
depend on the choice of origin, with this dependence arising through the $W$-term. 
In principle, the $A$-term and $K$-term are not directly affected by the origin: 
the $A$-term depends only on relative positions, which are invariant under a change of origin, 
and the $K$-term involves the phase, which is not a spatial quantity.
In contrast, the $W$-term depends on each individual's absolute position and is therefore 
sensitive to the chosen origin. This origin dependence directly influences the position 
dynamics in Eq.~\eqref{model}, which in turn affects the phase dynamics in Eq.~\eqref{model2}.

\subsection{\label{sm} Spiral current in the fixed-origin model}

\begin{figure}
\includegraphics[width=12cm]{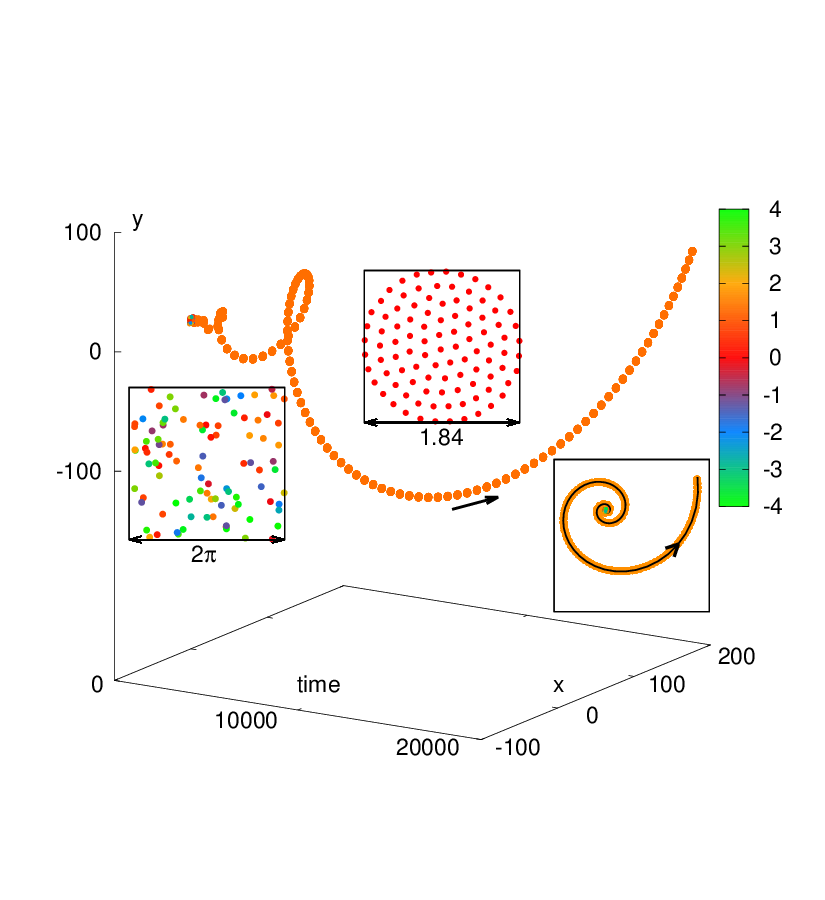}
\caption{(Color Online)
Outward spiral, the numerical time-trajectory of the cluster of swarmalators 
by Eqs.~\eqref{model} and \eqref{model2}.
Initially, at time $t=0$, swarmalators are randomly distributed over 
$2\pi\times 2\pi$ square (see the left inset) with random phases 
whose values are represented in the color bar right.
This initialization of positions and phases is used throughout this work 
unless otherwise specified.
The data points representing the swarmalators' positions 
in the spatiotemporal space are depicted at each discrete time steps $t=0,100,200,..\,$. 
Phase-sync of $\theta_{\rm s}=1.329784..=\theta_i$ for all $i$ is reached 
between $t=200$ and $t=400$ (see the colors become same). 
After synchronization, the trajectory of outward spiral appears.
The arrow is the direction of the outward motion.
The seeming circle each data thereafter is the cluster of all the swarmalators 
in system. The middle inset is the zoom-in of such cluster, which reveals 
that is a disc formed by the swarmalators.
The right inset is the projection of the spiral trajectory onto the $x$-$y$ plane.
The solid curve is the theoretical prediction given by Eq.~\eqref{rp}.
For the parameters, $A=0.01$, $W=0.04$, and $K=0.16$ are used, 
and the system size is $N=100$. 
	}
\label{fg-BSO}
\end{figure}

We first consider the case of a fixed origin, $\tilde O=O= (0,0)$.
In this case, the spatial angle simplifies to 
$\tilde\phi_i=\phi_i=\tan^{-1}y_i/x_i$.
Using this $\phi_i$, we perform numerical simulations for various 
values of $W$, $K$, and $N$, while keeping $A=0.01$.
As shown in Fig.~\ref{fg-BSO}, 
the swarmalators all in the system synchronize their phases
and form a cluster that exhibits the outward spiral motion.
%%%%%%%%%%%%%%%%%%%%%%%%%%%%%%%
%%%%%%%%%%%%%%%%%%%%%%%%%%%%%%%
This spiral behavior can be understood as follows.
In Refs.~\cite{Kevin17,Lee21,gLee21}, it has been known that the steady-state
pattern of the $A$-term is a disc of radius less than $1$, which is composed of 
all swarmalators in the system while synchronized.
The middle inset, a zoom-in of the data points from the spiral at $t=20\,000$, 
displays a pattern that looks like such a disc.
In the inset, the colors of data points are indistinguishable
as implying the phases are locked at a
synchronized value $\theta_{\rm s}$ given as
a fixed-point solution of Eq.~\eqref{model2}.

When the disc of the inset is stationary for the $A$-term in Eq.~\eqref{model}
and its radius is much smaller than $\|{\mathbf r}_i\|$ for all $i$, 
Eq.~\eqref{model} can be approximated as
\begin{equation}
\label{dap}
\dot{\mathbf r}
	= W\left(\hat x\cos(\phi +\theta_{\rm s})
	+ \hat y \sin(\phi +\theta_{\rm s})\right)\,,
\end{equation}
where ${\mathbf r}=(x,y)$ is the position average of the swarmalators 
and $\phi=\tan^{-1}(y/x)$. 
When the radial unit vector $\hat\rho$ and the angular (rotational) unit vector $\hat\phi$ 
are defined at the position ${\mathbf r}=(x,y)$,
Eq.~\eqref{dap} is rewritten by
\begin{equation}
	\label{rpc0}
\dot{\mathbf r}
=W\left(\hat\rho\cos\theta_{\rm s}
	+\hat\phi\sin\theta_{\rm s}\right)
\end{equation}
for $\rho\equiv\sqrt{x^2+y^2}$.
Since $\dot {\mathbf r}=\hat\rho\dot\rho+\hat\phi\dot\phi\rho$, comparing this with 
Eq.~\eqref{rpc0} yields 
\begin{eqnarray}
	\label{spu}
		\dot \rho &=& W\cos\theta_{\rm s}, \\
	\label{spu2}
		\rho\dot \phi &=& W\sin\theta_{\rm s}.
\end{eqnarray}
From Eq.~\eqref{spu}, we obtain 
\begin{align}
\frac{d}{dt}\ln \rho ={\dot \rho}/{\rho}\nonumber ={W\cos\theta_{\rm s}}/\rho.
\end{align}
Substituting this into Eq.~\eqref{spu2}, we find
\begin{align}
\frac{d\phi}{dt}= \tan\theta_{\rm s}\frac{d}{dt}\ln \rho,
\end{align}
which integrates to  
\begin{equation}
	\label{rp}
	\rho 
	= \rho_{\rm p} \exp\left(\frac{\phi-\phi_{\rm p}}
	{\tan\theta_{\rm s}}\right)\,,
\end{equation}
where $\rho_{\rm p}$ and $\phi_{\rm p}$ denote the polar coordinates of 
a reference point along the trajectory.
To verify the validity of Eq.~\eqref{rp},
we project the data displayed in Fig.~\ref{fg-BSO} onto the $x$-$y$ plane, 
so that obtain the right inset. The solid curve therein, 
which closely matches the numerical data, is plotted using Eq.~\eqref{rp}.
Notably, this curve corresponds to the well-known logarithmic spiral, a pattern 
that appears widely in nature-from biological structures to cosmic formations~\cite{lgspr}.

Since the spiral extends outward from the origin,
$\rho$ increases over time. According to Eq.~\eqref{spu}, 
$|\theta_{\rm s}|<\pi/2$ is required for the spiral current to emerge.
The same condition is consistent with the spiral behavior described 
by Eqs.~\eqref{spu2} and \eqref{rp}.
In our numerical simulations, this inequality always holds for the observed spirals. 
This suggests that qualitatively different states may arise 
when $|\theta_{\rm s}|> \pi/2$.

In the numerical data, the condition $|\theta_{\rm s}|> \pi/2$ is observed
when all the swarmalators converge toward the origin $(x,y)=(0,0)$.
After an initial transient period, the phases satisfy $|\theta_{i}|>\pi/2$ for 
all $i$, as $\theta_{i}\to \theta_{\rm s}$ over time.
In this regime, the radial component of the $W$-term points inward toward the 
origin (see Fig.~\ref{fg-model}), driving the swarmalators to cluster near the origin.
This state is characterized by the absence of collective motion or current.
Note that the $A$-term neither promotes nor resists this convergence, as it 
depends only on relative positions, which are unaffected by the choice of origin.

\begin{figure}
\includegraphics[width=9.0cm]{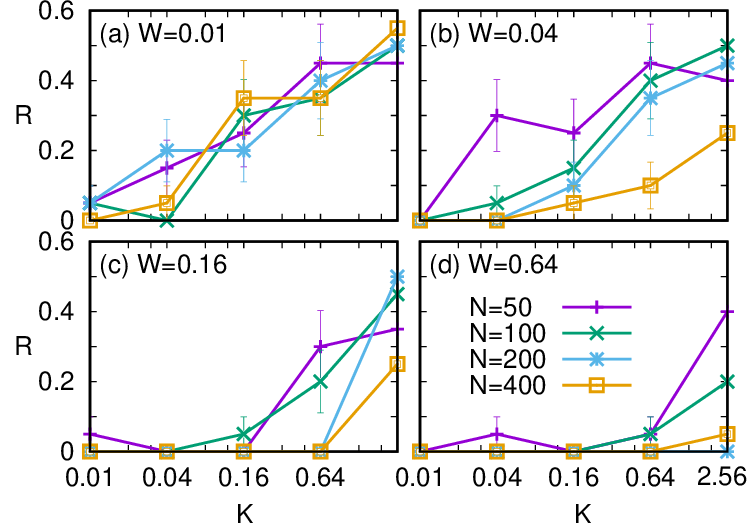}
\caption{(Color Online) 
	The ratio $R$ of spiral emergence is evaluated across various 
	combinations of $W$, $K$, and $N$. For each parameter set $(W,K,N)$, 
	whereby $W=0.01,0.04,0.16,0.64$, $K=0.01,0.04,0.16,0.64.2.56$, and $N=50,100,200,400$, 
	simulations are performed using 20 different initial configurations. 
	The resulting values of $R$ are plotted in panels (a)-(d). 
	The legend for $N$, shown in panel (d), applies to all panels.
	}
\label{fg-stat}
\end{figure}

We investigate the frequency of spiral current formation by examining 
the ratio $R=N_{\rm c}/N_{\rm a}$,   
where $N_{\rm c}$ is the number of appearance of the spiral current, 
and $N_{\rm a}$ is the number of all the tested configurations
each sampled from the uniform distribution 
described in the caption of Fig.~\ref{fg-BSO}.
The $R$ is measured for various combinations of $W$, $K$, and $N$. 
The results are presented in Fig.~\ref{fg-stat}.
%%%%%%%%%%%%%%%%%%%%%%%%%%%%%%%
%%%%%%%%%%%%%%%%%%%%%%%%%%%%%%%
Overall, the results show that $R$ tends to increase with $K$ but 
decrease with $W$. The non-decreasing trend of $R$ with respect to $N$, 
observed in Fig.~\ref{fg-stat}(a) for smaller $W$, 
does not persist in panels (b)-(d) for larger $W$.
Across all panels in Fig.~\ref{fg-stat}, $R$ remains below $0.5$ in most cases,
indicating a general tendency toward no-current states.
Since the initial positions are uniformly distributed within a square centered 
at the origin, the $A$-term alone without the influence of the $W$-term naturally 
forms a stationary disc at the origin. 
The disc formation is compatible with inward motion driven by 
$|\theta_{\rm s}|>\pi/2$, but not with outward motion when $|\theta_{\rm s}|<\pi/2$.
We suggest that this asymmetry may explain the observed bias toward no-current states, 
as reflected by the overall trend of $R<0.5$.

Direct manipulating the phase variables can trigger the emergence of spiral currents. 
Even pinning the phase of just one swarmalator is enough.  
In our observations (not shown here), a spiral current forms from 
a previously no-current state clustered around the origin simply by 
fixing the phase of a single swarmalator, for example at $\pi/4$, which lies   
within the interval $(-\pi/2,\pi/2)$. 
According to Eq.~\eqref{model2}, this pinned phases sets the synchronized 
phase $\theta_{\rm s}$, causing the swarmalators to move outward from the origin. 
This outward motion, combined with the disc-clustering dynamics, results in the 
formation of a spiral current.

The spiral current, as described by Eq.~\eqref{rp}, 
leads to an unbounded trajectory: the spiral radius grows exponentially 
over time, eventually diverging to infinity. 
As a result, the practical relevance of such a current is limited. 
Since swarmalator models are often motivated by real-world applications
such as controlling autonomous drones or modeling biological collectives,
it is more meaningful to focus on models whose 
results are confined to a finite spatial domain. 
In the following subsection, we demonstrate that by redefining the origin 
as the position center of the swarmalators, 
Eqs.~\eqref{model} and \eqref{model2} 
can give rise to a collective current that remains spatially bounded, 
avoiding the issue of divergence.

\subsection{\label{dcs}
Circular current in the dynamic origin model}

To prevent the system from diverging into infinite space, 
we redefine the origin as the position center of the swarmalators, 
$\tilde O=O_{\rm PC}=\sum_i {\mathbf r}_i/N$.
Before going further, we remark that this kind of movable origin composed of the 
swarmalators' positions is not that artificial compared to the fixed one.
It is hard to consider biological agents could share the fixed origin that is in fact 
highly ideal and abstract. 
In this sense, an origin as the reflection of the interacting constituents is rather practical. 
We believe $O_{\rm PC}$ can be a representative of such origins.

In case of $O_{\rm PC}$, 
even when the system is in sync phase 
($\theta_{\rm s}=\theta_i$ for all $i$), 
each swarmalator in a cluster (if formed) experiences 
a different direction of driving.
This arises because the spatial angle  
$\tan^{-1}(\tilde y_i/\tilde x_i)$ is added to the common phase angle 
$\theta_{\rm s}$. As a result,
a coherent motion of the clustered swarmalators with respect to
the origin, which can drive out the cluster far from the origin, 
is no longer allowed.
Individual divergence of swarmalators in separate directions is 
prohibited by the linear attraction in $A$-term.

When $|\theta_{\rm s}|>\pi/2$, the inward driving readily compresses
the disc expected by $A$-term, and this results in no-current state of 
swarmalators gathering around the origin $O_{\rm PC}$.
We numerically confirmed this expectation (not shown here). 
Compared to this, the situation is not that 
simple if $|\theta_{\rm s}|<\pi/2$.

\begin{figure}
\includegraphics[width=9.0cm]{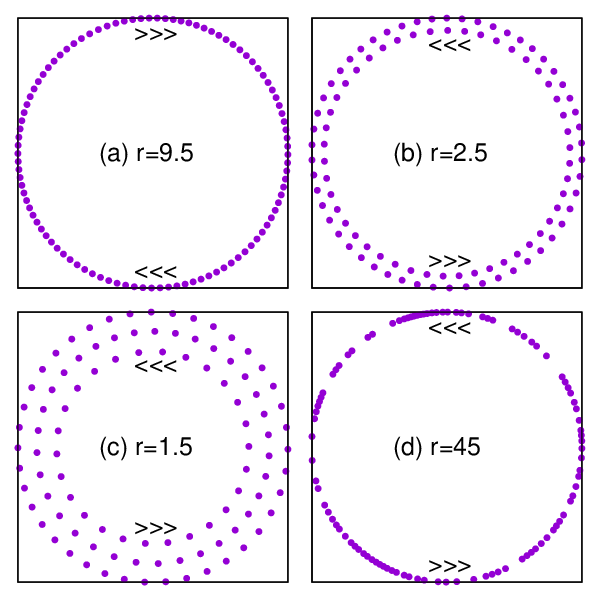}
\caption{(Color Online) Snapshots of circular currents appearing for the model
        with position-center origin $O_{\rm PC}$ (see text).
        Each panel shows the distribution of swarmalators 
	at a time long after transient period
	(the data points represent swarmalators). 
        At the center of panel, 
	the radius of circular pattern is written.
	The rotation direction is indicated with the arrow heads 
	($<<<$ or $>>>$).
        The pattern in each panel is composed of $N=100$ swarmalators.
        (a) shows the (single-)ring formed when $W=0.16$ and $K=1.28$.
        The double-ring in (b) appears when $W=0.02$ and $K=1.28$.
        (c) is the triple-ring at $W=0.01$ and $K=1.28$.
        The ring of large radius and of inhomogeneous spacing 
	in (d) appears when $W=0.64$ and $K=2.56$.
	}
\label{fg-ms}
\end{figure}

Figure~\ref{fg-ms}
shows a few results accompanied with $|\theta_{\rm s}|<\pi/2$. 
%%%%%%%%%%%%%%%%%%%%%%%%%%%%%%%
%%%%%%%%%%%%%%%%%%%%%%%%%%%%%%%
Each point in the panels represents an individual swarmalator.
The circular patterns therein are not time trajectories 
but distributions of swarmalators in their own time snapshots.
Figure~\ref{fg-ms}(a)-(c) display single- or multi-ring patterns
emerge. As those patterns are rotationally symmetric, 
the radial components of the driving $W$-term are balanced.
The rotational components lead to collective circular current 
of the circularly distributed swarmalators.
According to numerical observations,
each one looks not a long-lived transient but an attractor.

For a simple argument of the stability of the circular currents,
we have examined linear stability as follows.
Let $\{{\mathbf r}_i\}$ be the collection of the positions 
of the swarmalators that form a ring of radius $r$.
To make the calculation simple, we choose the direction of 
$\tilde{\mathbf r}_i$ for a certain $i$ as $x$-axis so that it holds
$(\tilde x_i=r,\tilde y_i=0) + O_{\rm PC} = (x_i,y_i)$.
Consider ${\mathbf r}_i + \delta = (x_i+\delta_x,y_i+\delta_y)
=(r+\delta_x,\delta_y)+O_{\rm PC}$
for small $\|\delta\| \equiv \sqrt{\delta_x^2+\delta_y^2}$. 
When the pattern is circularly symmetric, 
the center does not move, i.e., $\dot O_{\rm PC}=0$
in that i) all vectors by $W$-term are summed out
and ii) the conservative $A$-term generically cannot move
the center. Plugging ${\mathbf r}_i + \delta$ into Eq.~\eqref{model},
and linearizing it for $\delta$, one can obtain 
in use of $\dot O_{\rm PC}=0$ that 
\begin{eqnarray}
	\label{lineq}
	\dot\delta_x &=& 
	-\frac{A}{N}\sum_j\left\{
	\left(\frac{1}
	{\|{\mathbf r}_j-{\mathbf r}_i\|^2}
	+\frac{2\Delta x_{j,i}^2}
	{\|{\mathbf r}_j-{\mathbf r}_i\|^4} 
	\right)\delta_x
	+\frac{2\Delta x_{j,i}\,\tilde y_j
	}
	{\|{\mathbf r}_j-{\mathbf r}_i\|^4}\delta_y\right\}
-\left(\frac{W}{r}\sin\theta_{\rm s}\right)\delta_y
\nonumber\\
	\dot\delta_y &=& 
	-\frac{A}{N}\sum_j\left\{
	\left(\frac{1}{\|{\mathbf r}_j-{\mathbf r}_i\|^2}
	+\frac{2
	\tilde y_j^2
	}{\|{\mathbf r}_j-{\mathbf r}_i\|^4} 
	\right)\delta_y
	+\frac{2\Delta x_{j,i}\,\tilde y_j
	}
	{\|{\mathbf r}_j-{\mathbf r}_i\|^4}\delta_x\right\}
+\left(\frac{W}{r}\cos\theta_{\rm s}\right)\delta_y
\,,
\end{eqnarray}
where $\Delta x_{j,i}\equiv x_j-x_i$.

For rotational symmetric pattern, one knows that
$\sum_j \Delta x_{j,i}\,\tilde y_j/
\|{\mathbf r}_j-{\mathbf r}_i\|^4=0$ because
there is a permutation $\sigma$ for which
$\tilde y_j=-\tilde y_{\sigma(j)}$ and 
$\Delta x_{j,i}=\Delta x_{\sigma(j),i}$. 
Then, Eq.~\eqref{lineq} is rewritten as
$\dot \delta = -M \delta$ with
\begin{equation}
	\label{M}
	M=
\begin{pmatrix}
\frac{A}{N} \sum_j \left(\frac{1}
         {\|{\mathbf r}_j-{\mathbf r}_i\|^2}
         +\frac{2\Delta x_{j,i}^2}
         {\|{\mathbf r}_j-{\mathbf r}_i\|^4} 
         \right) 
	& \frac{W}{r}\sin\theta_{\rm s} \\
0 & 
	 \frac{A}{N} \sum_j \left(\frac{1}
          {\|{\mathbf r}_j-{\mathbf r}_i\|^2}
          +\frac{2 \tilde y_j^2}
          {\|{\mathbf r}_j-{\mathbf r}_i\|^4}
          \right)
	  -\frac{W}{r}\cos\theta_{\rm s}
\end{pmatrix}\,,
\end{equation}
whose eigenvalues are $\lambda_1 = M_{1,1}$ and $\lambda_2 = M_{2,2}$. 
We note $\lim_{N\to\infty}
\sum_j \|{\mathbf r}_j-{\mathbf r}_i\|^{-2}/N
= \frac{1}{r^2}\int_{0}^{\pi}d\varphi\frac{1}{1-\cos\varphi}\to\infty$. 
So that $\lambda_1$ and $\lambda_2$ become positive 
if $N$ is larger than a certain number.
Positive $\lambda_1$ and $\lambda_2$ guarantee that
a perturbation whose effect is considered in the leading order
decreases, and then finally disappears.
Our argument for linear stability is also valid for the multi-rings.
Hence, it is expected that 
the patterns displayed in Fig.~\ref{fg-ms}(a)-(c) are linearly stable.

Interestingly, the pattern shown in Fig.~\ref{fg-ms}(d) is not 
that symmetric:
though this looks circular, inter-swarmalator spacings are not uniform.
Still, it is neither deforming nor moving to somewhere.
Thus the circular pattern is more or less maintained. 
The circular shape is perceived by the seemingly fixed radius 
in sprite of the non-uniform spacing between swarmalators. 
In this observation, one may consider that the radial stability is stronger 
than the rotational counterpart. 
The $x$-direction we set above for stability analysis
can be regarded as radial direction
and the eigenvector of $\lambda_1$ is $(1,0)$.
The perpendicular $y$-direction heads to rotational direction, 
and the eigenvector of $\lambda_2$ has $y$-component.
When $r$ is large like in Fig.~\ref{fg-ms}(d), one may see
$\lambda_2 \approx \lambda_1 - (W/r)\cos\theta_s < \lambda_1$ 
for $|\theta_{\rm s}|<\pi/2$ neglecting $O(1/r^4)$ terms, which
implies the radial direction alone can remain linearly stable.
This argument is consistent with the observation 
that the radius is kept while the spacing is not.

\begin{figure}
\includegraphics[width=9.5cm]{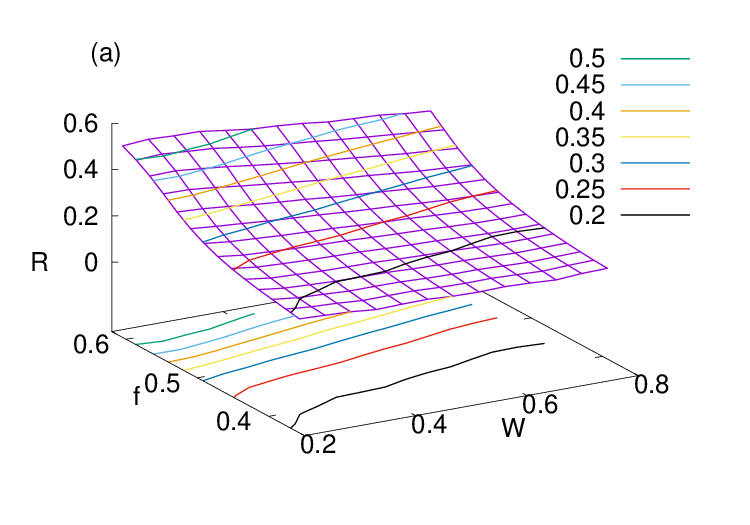}
\includegraphics[width=6.5cm]{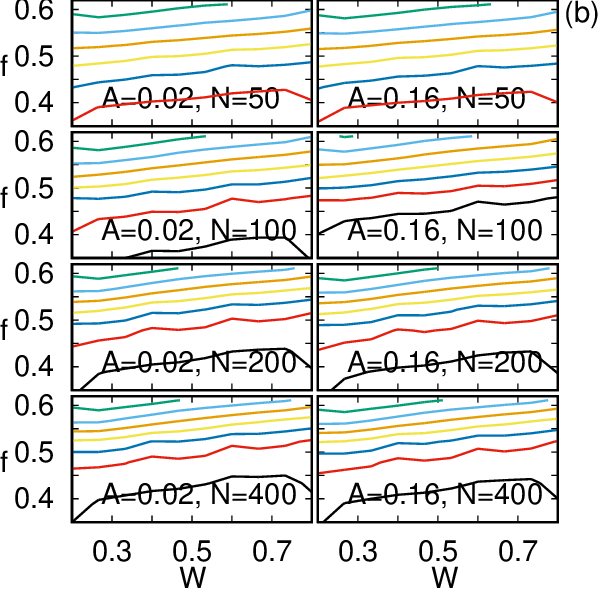}
	\caption{(Color Online) Surface of $R$ on $W$-$f$ plane 
	obtained for $A=0.08$ and $N=400$ in (a);
	$R$ is the ratio of emergence of circular currents 
	and $f$ is a phase-initialization parameter 
	(see text for the detail of $f$).
	The curves both on surface and on bottom are the contours of $R$.
	For (a), we have scanned $W$-$f$ plane on the 
	$0.2 \le W\le 0.8$ and $0.3 \le f\le 0.7$ region with
	$\Delta W=\Delta f=0.02$ step, i.e.,
	$21\times31$ number of $W$ and $f$ pairs are examined~\cite{note1}.
	For each pair of $W$ and $f$, 200 random samples are tested.
	More contours of $R$
	are on the $W$-$f$ panels in (b), of which colors represent 
	the same values of $R$ listed at the upper-right corner of (a).
	For every panel, the $W$-$f$ plane is scanned in the same way
	done for (a). The used $A$ and $N$ is written in each panel.
	}
\label{fg-est}
\end{figure}

We have performed the intensive numerical test for the ratio $R$
of the emergence of the circular currents with 
$|\theta_{\rm s}|<\pi/2$.
For this, an initialization-parameter $f$ is introduced,
which controls the fraction of initial phases in 
the interval $(-\pi/2,\pi/2)$.
The initialization explained in the caption of Fig.~\ref{fg-BSO}
corresponds to the $f=0.5$ case.
The control parameter $f$ is introduced in the expectation that
the larger $f$ may promote the more chance of 
$|\theta_{\rm s}|<\pi/2$. 
The results are summarized in Fig.~\ref{fg-est}.
%%%%%%%%%%%%%%%%%%%%%%%%%%%%%%%
%%%%%%%%%%%%%%%%%%%%%%%%%%%%%%%

With numerous numerical data, we obtain $R$ on $W$-$f$ plains. 
This is done for several $A$ and $N$ while fixing $K=1$.
Figure~\ref{fg-est}(a) displays the $R$-surface 
with several contours both on surface and bottom
(see the caption for details).
Therein, $R$ decreases for $W$ like in Fig.~\ref{fg-stat} while
increases for $f$ as expected.
The surface and contours shown in Fig.~\ref{fg-est}(a) 
are the typical ones repeatedly observed in our numerical study.
Its shape does not substantially change in our test 
for $A=0.02,0.04,0.08,0.16$ and $N=50,100,200,400$.
Some of the results are displayed in Fig.~\ref{fg-est}(b).

Although $R$ increases with $f$, 
a bias toward no-current states ($R<0.5$ on average) persists, 
at least up to $f=0.6$. 
Unlike the inward driving by $|\theta_{\rm s}|>\pi/2$,
the outward driving by $|\theta_{\rm s}|<\pi/2$
should make a balance with the inward tendency of $A$-term 
that favors disc formation,
by reaching a compromise like the patterns in Fig.~\ref{fg-ms}.
If the balance fails, the driving is already meant to change to 
inward one necessarily accompanied with $|\theta_{\rm s}|>\pi/2$.
We consider this could be a reason for the bias toward no-current states.

The response for $N$ can be found 
in the columns of the panels in Fig.~\ref{fg-est}. 
It looks that the same colored contours shift upward
and the spacing between them decreases, as $N$ increases.
This observation suggests
the emergence of the collective circular currents becomes of less chance
for larger $N$.
We leave the study about the thermodynamic property
as future works, which will be interesting  
with local interactions instead of the global ones 
inherited from the standard swarmalator model.
We finally remark 
that controlling/managing the circular currents via manipulating 
the phase variable is also possible, as mentioned 
at the end of Sec.~\ref{sm} for the spiral currents.

\section{\label{sum}Summary}

We have proposed a swarmalator model that generates two distinct types of 
collective currents, depending on the choice of origin. 
Specifically, we introduced a simple non-conservative driving term that depends on 
both the phase and the origin, replacing the standard phase-dependent terms in the 
spatial dynamics of the original swarmalator model. 
The behavior of the system under this modification was then explored numerically.

When the origin is fixed, the model produces a collective current in the form of spiral, 
with the swarmalators forming a phase-synchronized cluster. 
The resulting trajectories align closely with a logarithmic spiral, as 
confirmed by numerical simulations.
However, the exponential growth of the spiral's radius leads to unbounded motion,
which is impractical in real-world scenarios.

To address this, we redefined the origin dynamically as the center of 
swarmalators' positions. 
In this co-moving frame, the effective driving term changes, 
leading to a different form of collective currents 
of circular patterns within finite space. 
This bounded, finite-size dynamical state offers more practical relevance for 
applications of swarmalator models, such as in engineering and biological systems.

In both cases, the system can also settle into a no-current state, where the 
swarmalators gather near the origin while remaining phase-synchronized. 
We investigated the occurrence rate of current-generating states 
across different model parameters and system sizes, finding a bias 
toward no-current states when the initial configurations are uniformly randomized. 
The emergence of current states can be encouraged by 
tuning the initial phase distribution.

Exploring the thermodynamic properties of these collective current states, 
especially under local interaction rules, remains an open direction for future research. 
Beyond simply promoting current states, the potential to actively manage 
or control them through targeted manipulation of phase variables presents 
a practically valuable avenue.

\section{Acknowledgments}
This research was supported by the National Research Foundation of Korea (NRF) 
grant funded by the Korea government (MSIT) (Grant No.~RS-2023-00276248 (H.K.L.) 
and RS-2024-00348768 (H.H.)).

\def\tb{\textbackslash}
 

\begin{references}
\bibitem{note1} The results in the $f>0.62$ or $f<0.35$ region are respectively
	sudden changes to their own extremes, i.e., to $R=1$ or $R=0$,
	and then stay at the extreme values. As this generates numerically
	unstable contours, we do not include them in Fig.~\ref{fg-est}.
\end{references}
\end{document}